\documentclass[twocolumn,showpacs,preprintnumbers,amsmath,amssymb]{revtex4}
\usepackage{graphicx}
\usepackage{dcolumn}
\usepackage{bm}
\usepackage{multirow}
\def\bTet{{\bf Tetra}}
\def\bOrt{{\bf Ortho}}
\def\iTet{{\it Tetra}}
\def\iOrt{{\it Ortho}}

\begin{document}

\preprint{}

\title{Prediction of stable insulating intermetallic compounds}

\author{M. Mihalkovi\v{c}, M. Kraj\v{c}\'{\i}}
\affiliation{Institute of Physics, Slovak Academy of Sciences, 84511 Bratislava, Slovakia}
\author{M. Widom}
\affiliation{
Department of Physics, Carnegie Mellon University\\
Pittsburgh, PA  15213
}

\date{\today}
\begin{abstract}
We explore the stability of structure exhibiting hybridization gaps across a broad range of binary and ternary intermetallic compositions by means of band structure and total energy calculations. This search reveals previously unknown metal-based insulators, some with large gaps exceeding 1 eV, such as Al$_2$Fe and Al$_4$IrRe.  We confirm large gaps using a hybrid density functional including exact exchange, and predict a gap of 2.2 eV for AlMnSi in the Pearson type tP6 structure, which is a chemically ordered ternary variant of the prototype MoSi$_2$ (Pearson type tI6) structure.
\end{abstract}

\maketitle

\section{Introduction}

Certain intermetallic alloys are electrically insulating despite the
good metallic characters of their constituent elements.  A classic
example of this is provided by the compound Al$_2$Ru which is
insulating as result of gaps in its density of states around its Fermi
level~\cite{nguyen1992}.  Such materials pose intriguing questions as
to the origins of their insulating properties and additionally may
find potential applications, for example as thermoelectric materials~\cite{Takeuchi10}.

Previous studies
~\cite{Weinert98,Krajci02a,Krajci02b,Fredrickson04a,Fredrickson04b}
have explained the origins of this gap, and generalized the families of
compounds in which it is expected to occur.  The essential mechanism
is to establish a gap between bonding and antibonding states through
the creation and occupation of hybridized orbitals, then to exactly
fill the bonding states.  For structures based on the prototype
MoSi$_2$ (Pearson type tI6) and TiSi$_2$ (Pearson type oF24) this occurs at
composition (Al)$_2$(Fe), where the parenthesis around (Al) indicates
an aluminum-group metal of valence three (specifically Al or Ga) and
(Fe) is an iron-group transition metal of valence eight (i.e. Fe, Ru
or Os).  Notice these choices result in a total of fourteen valence
electrons per transition metal atom.

Gap width may be controlled by selection of the specific combinations
of elements, or by alloying with adjacent elements in the periodic
table (e.g. replacing the (Fe)-group element with a combination of
(Mn)-group and (Co)-group), while maintaining constant valence per
transition metal atom.  Alternatively, we may replace half of the
(Al)-group elements with (Si)-group, while simultaneously replacing
the (Fe)-group transition metal with a (Mn)-group element.

Here we report a thorough study of the tI6 and oF24 alloy families.
The crucial new feature of our work is we simultaneously evaluate the
compounds' enthalpies of formation, so that we may predict cases where
the proposed structures are likely to form as stable compounds, as
bandgap engineering through chemical substitution requires monitoring
both variation of band gaps as well as thermodynamic stability
relative to competing phase formation~\cite{Sun11}.  In this manner,
we predict the occurrence of thirteen previously unknown compounds,
each of which exhibit electronic band gaps.  We then checked the gaps
predicted by conventional DFT utilizing a hybrid functional including
exact exchange.  Among our newly predicted stable compounds is Al-Fe,
with a gap of 1.1 eV, Al-Ir-Re, with a gap of 1.3 eV, and Al-Mn-Si
with a gap of 2.2 eV.

\section{Structures}

\begin{figure}
\includegraphics[width=2.75in]{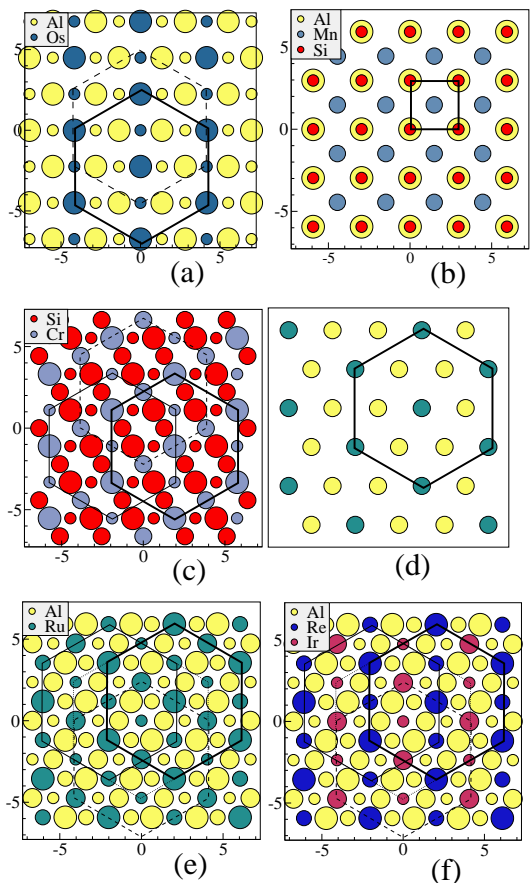}
\caption{\label{fig:struct} Structures of the class. Colors indicate atomic species, size indicates vertical height. (a) Al$_2$Os in the Pearson tI6 (MoSi$_2$) structure. TM hexagons in alternate layers outlined in solid and dashed lines. (b) AlMnSi in tP6 structure. View along (001) axis reveals uniform chemical identity of layers, and relationship to cP2 (CsCl) structure. (c) CrSi$_2$ in the hP9 structure revealing three stacked hexagons. (d) Generic hexagon motif in single layer. (e) Al$_2$Ru in oF24 (Si$_2$Ti) structure reveals four stacked hexagon motifs. (f) Ternary decoration of Al$_4$IrRe in oF24 structure.}
\end{figure}

Al$_2$Ru has been reported to occur in both the tI6~\cite{Obrowski63}
and the oF24~\cite{Edshammar66} structures.  Indeed, both structures
share common structural elements, and are stabilized by similar
mechanisms.  The essential structural feature is an Ru-centered ring
of six Al atoms (see Fig.~\ref{fig:struct}), repeated periodically to
tile the plane.  A third structure, based on the CrSi$_2$ prototype (hP9)
shares a similar motif.  The ring is a slightly irregular hexagon,
with only a 2--fold rotational symmetry.  This ring lies
perpendicular to the [110] axis in tI6 and the [001] axis in hP9 and
oF24.  The differences between structures rests in the stacking of
these TM-centered rings (TM stands for transition metal).  Consider a
triangle of three TM atoms within a layer.  There are four possible
sites for the TM in an adjacent layer, either the vertex site or else
one of the three edge centers of this triangle.  In tI6 the TM
alternates between two such sites, in hP9 between three and in oF24
all four sites are utilized.

Chemical ordering in ternary compounds lowers the symmetry from its binary prototype.  For example, the MoSi$_2$ prototype is Pearson type tI6, space group I4/mmm, while the optimal decoration for AlMnSi is Pearson type tP6 (Primitive, not body centered) with space group P4/nmm . The chemical occupation is uniform within layers when the structure is viewed down the 4--fold axis.  For example, an Al$_2$Mn layer alternates with an Si$_2$Mn layer (see Fig.~\ref{fig:struct}$b$), rather than mixing Al and Si within a layer. 
Likewise the Si$_2$Ti prototype of Pearson type oF24 has space group Fddd while the optimal decoration for AlIrRe maintains Pearson type oF24 but lowers the space group to F222 (Fig.\ref{fig:struct}$f$). Energetically optimized Wyckoff positions for the two prototypical structures are given in Table~\ref{tab:struc}.

\section{Methods}

We utilize VASP~\cite{Kresse93,Kresse96} to carry out first principles total energy calculations.  All atomic positions and lattice parameters are fully relaxed.  We increase our k-point densities until energies have converged to 1 meV/atom.  Default energy cutoffs are employed, which are sufficient to converge energy differences to this tolerance. Because of the volume relaxation, our energy may be considered as an enthalpy at P=0 and T=0K.  We adopt projector augmented wave potentials~\cite{Blochl94,Kresse99}, and for a density functional we choose the PBE generalized gradient approximation.  Spin polarization is utilized in appropriate cases.

Conventional density functional theory generally underestimates band gaps, so we employ a hybrid functional~\cite{Paier05}, HSE06~\cite{Heyd06}, that includes a portion of exact exchange, to obtain more reliable values.  However, the hybrid functional is computationally expensive so we do this only in selected cases. We also report pseudogap widths, which we define as the energy interval within which the density of states remains below 0.20 states/ev/atom.

Our stability calculations follow methods outlined in~\cite{Mihal04}.  Primarily, we calculate enthalpy of formation $\Delta H$ by subtracting the calculated enthalpies from the enthalpies of the pure elements in their stable forms.  In addition to the enthalpies of our tI6 and oF24 structures, we include all known structures within the given alloy system as well as selected hypothetical structures that occur in chemically similar systems.  

In the case of ternary systems, the atomic decoration is not
constrained, due to the mixing of TM species, or substituting
(Si)-group for (Al)-group elements.  To search for optimal patterns of
chemical ordering, we first exhaustively enumerate all
symmetry-independent configurations within cells specified below.  We
examine for two representative alloy systems of each class, to check
for consistency of the optimal chemical decoration, and then
subsequently apply the optimal decoration to all other ternary alloy
systems from the corresponding family.

{\it (B/C)Mn class.} Our representative alloy systems were AlMnSi and
AlReSi.
In the MoSi$_2$.tI6 prototype structure, we limited our ground state search to
1$\times$1$\times$2 and $\sqrt{2}\times\sqrt{2}\times$1 supercells.
Each of these had 10 independent configurations.
In the Si$_2$Ti.oF24 prototype case, the unit cell with 8 Si and
8 Al atoms occupying "Si" sites yields 293 independent configurations.  We limited our ground-state search to configurations of relatively high symmetry.

{\it Al(Mn/Co) class.} Our representative alloy systems were AlIrRe
and AlReRh.  In the MoSi$_2$.tI6
prototype structure, there are 2 independent configurations each in the
1$\times$1$\times$2 and $\sqrt{2}\times\sqrt{2}\times$1 supercells.  In the Si$_2$Ti.oF24 prototype case, there were 8 independent
assignments for transition metals on the ``Ti'' sites. 

Given a database of relaxed structures, we calculate the convex hull of enthalpy as a function of composition.  Predicted stable structures occupy the vertices of the convex hull.  For structures that are {\em unstable}, we report the energy $\Delta E$ by which they lie above the convex hull. For stable compounds, negative values of $\Delta E$ indicate the formation enthalpy with respect to compounds of adjacent compositions.

\begin{figure}
\includegraphics[width=2.75in,angle=-90]{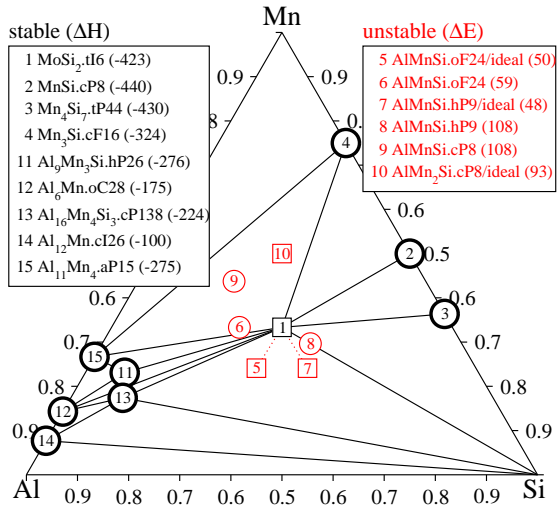}
\caption{\label{fig:al-mn-si} (color online) Phase diagram of Al-Mn-Si. Heavy circles indicate known stable phases, light circles indicate known high temperature phases, squares indicate unknown structures or compositions. $\Delta H$ values are enthalpy of formation (meV/atom) of stable structures. Positive $\Delta E$ values are energies above the convex hull for unstable structures, while negative $\Delta E$ is the formation enthalpy with respect to adjacent compositions for stable structures.}
\end{figure}

Fig.~\ref{fig:al-mn-si} gives an example for the case of Al-Mn-Si.  We
reproduce the stability~\cite{Prince93} of most known low temperature
binaries, and two of the ternaries. One of these ternaries
(Al$_{16}$Mn$_4$Si$_3$.cP138) is an icosahedral quasicrystal
approximant, and exhibits a gap in its density of states. However, two
reported phases (Al$_{0.6}$Mn$_{0.8}$Si$_{0.4}$ and
Al$_{8}$Mn$_{3}$Si$_{9}$) have unknown structure and hence are omitted
from our study, and three reported phases (Pearson types cP8, oF24 and
hP9) exhibit partial site occupancy or mixed chemical occupancy and
lie off their ideal stoichiometries, and hence are presumed to be
stable only at high temperatures.  Meanwhile, our predicted lowest
energy of AlMnSi in the tP6 structure has not been reported
experimentally.  We suspect the prediction is correct, as the $\Delta
E=-48$ meV/atom is well beyond the expected uncertainty of DFT, and
our exact exchange calculation confirms stability of tP6.  Perhaps
oF24 is a high temperature phase whose entropy inhibits formation of
the true low temperature tP6 phase through conventional sample
preparation methods.

\begin{table}
\begin{tabular}{c|cccc||c|cccc}
\hline
     &    &  $x$    &   $y$  &    $z$     &      &   & $x$   &   $y$      & $z$  \\
   Al&2c  & 0.75  & 0.75 &  0.9014  &    Ir&4a & 0.0  &  0.0     &0.0 \\
   Mn&2c  & 0.75  & 0.75 &  0.2736  &    Re&4d & 0.75 &  0.75    &0.75\\
   Si&2c  & 0.75  & 0.75 &  0.5846  &   Al1&8f & 0.50 &  0.6706  &0.50\\
     &    &       &      &          &   Al2&8i & 0.25 &  0.0817  &0.25\\
\hline
\end{tabular}
\caption{\label{tab:struc}
Atomic structures of AlMnSi.tP6 (left block; space group P4/nmm (\#129), $a$=3.027\AA, $c$=7.932\AA) and Al$_4$ReIr.oF24 (right block; group F222 (\#22), $a$=4.745\AA, $b$=8.139\AA, $c$=8.925\AA). Wyckoff sites are labeled by site multiplicities.
}
\end{table}

\section{Results}

Table~\ref{tab:data} presents our predicted stable structures along with their band gaps (if any) and pseudogap widths. Complete data is given in Table~\ref{tab:data}. We confirm the stability of Al2Ru, Al2Os, Ga2Ru and Ga2Os in the oF24 structure, and AlGeRe in the tI6 structure. We predict the stability of (Al,Ga)4TcRh, (Al,Ga)4ReIr and AlGeMn in the oF24 structure and Al2Fe, AlSiMn, (Al,Ga)SiTc, GaSiRe, (Al,Ga)GeTc and GaGeRe in the tI6 structure.

We previously discussed the case of Al2Fe~\cite{Mihal12}.  Although total energy calculations predict it to be stable, the observed Al$_2$Fe phase is intrinsically disordered with low symmetry.  This phase may be thermodynamically stabilized at high temperature by its vibrational entropy.  It is probable that formation of the true low temperature stable phase is kinetically hindered at low temperatures.  The most interesting new prediction may be AlMnSi, as it is comprised of common elements and it has the largest predicted true gap (2.2 eV, 1 eV larger than elemental silicon)

\begin{table*}[h]
\begin{tabular}{ll|rrrrrr||ll|rrrrrr}
\hline
\hline
\multicolumn{2}{c|}{Al$_2$(Fe)} & \multicolumn{2}{c}{ } &
\multicolumn{2}{c}{PBE} & \multicolumn{2}{c||}{HSE06} &
\multicolumn{2}{c|}{Ga$_2$(Fe)} & \multicolumn{2}{c}{ } &
\multicolumn{2}{c}{PBE} & \multicolumn{2}{c}{HSE06} \\
\hline
Chem & Struct & $\Delta E$ & $\Delta H$ & $E_g$ & $E_p$ & $E_g$ & $E_p$ &
Chem & Struct & $\Delta E$ & $\Delta H$ & $E_g$ & $E_p$ & $E_g$ & $E_p$\\
\hline
\multirow{2}{*}{Al-Fe}
 & Ortho &  22 &-335 & 0.220 & 0.650& 0.739 & 1.287 &
\multirow{2}{*}{Ga-Fe}
 & Ortho & 127 & -94 & 0.141 & 0.469& 0.614 & 1.115 \\
 & \bTet & -21 &-358 & 0.008 & 0.401& 1.115 & 1.663 &
 & Tetra & 182 & -39 & 0.011 & 0.016& 0.001 & 0.016 \\
\hline                                                    
\multirow{2}{*}{Al-Ru}
 & \iOrt & -17 &-685 & 0.112 & 1.119 &  0.591 & 1.669 &
\multirow{2}{*}{Ga-Ru}
 & \iOrt & -20 &-377 & 0.107 & 0.928 &  0.559 & 1.448 \\
 & Tetra &  33 &-653 & 0.005 & 0.556 &  0.489 & 1.319 &
 & Tetra &  74 &-303 & 0.014 & 0.527 &  0.274 & 1.149 \\
\hline                                                    
\multirow{2}{*}{Al-Os}
 & \iOrt &-0.7 &-560 & 0.698&1.287& 1.061&  2.060 &
\multirow{2}{*}{Ga-Os}
 & \iOrt &  76 &-156 & 0.644&1.142& 1.199&  1.767 \\
 & Tetra & 0.7 &-559 & 0.354&1.151& 1.057&  1.882 &
 & Tetra & 107 &-125 & 0.008&0.987& 0.669&  1.752 \\
\hline
\hline
\multicolumn{2}{c|}{Al$_4$(Mn)(Co)} & \multicolumn{2}{c}{ } &
\multicolumn{2}{c}{PBE} & \multicolumn{2}{c||}{HSE06} &
\multicolumn{2}{c|}{Ga$_4$(Mn)(Co)} & \multicolumn{2}{c}{ } &
\multicolumn{2}{c}{PBE} & \multicolumn{2}{c}{HSE06} \\
\hline
Chem & Struct & $\Delta E$ & $\Delta H$ & $E_g$ & $E_p$ & $E_g$ & $E_p$ &
Chem & Struct & $\Delta E$ & $\Delta H$ & $E_g$ & $E_p$ & $E_g$ & $E_p$\\
\hline
\multirow{2}{*}{Al-CoMn}
&  Ortho &  43 & -364 &  0.274&  0.700& 0.690 & 1.413  &
\multirow{2}{*}{Ga-CoMn}                  
&  Ortho &  93 & -144 &  0.277&  0.554&     x &     x  \\
&  Tetra &  39 & -368 &  0.028&  0.394& 0.605 & 1.545  &
&  Tetra & 163 &  -74 &  0.014&  0.241&     x &     x  \\
\hline
\multirow{2}{*}{Al-TcRh}
& \bOrt  & -50 & -693 &  0.425&  1.125&     x &     x  &
\multirow{2}{*}{Ga-TcRh}                  
& \bOrt  & -76 & -402 &  0.307&  1.024& 0.805 & 1.666  \\
& Tetra  &  50 & -643 &  0.013&  0.528& 0.023 & 1.199  &
& Tetra  &  76 & -326 &  0.014&  0.429&     x &     x  \\
\hline
\multirow{2}{*}{Al-ReIr}
& \bOrt  & -40 & -597 &  0.702&  1.326& 1.260 & 1.999  &
\multirow{2}{*}{Ga-ReIr}                  
& \bOrt  & -18 & -234 &  0.905&  1.250& 1.426 & 1.833  \\
& Tetra  &  85 & -513 &  0.000&  0.009& 0.001 & 0.522  &
& Tetra  &  57 & -177 &  0.009&  0.751&     x &     x  \\
\hline
\hline
\multicolumn{2}{c|}{Al(Si)(Mn)} & \multicolumn{2}{c}{ } &
\multicolumn{2}{c}{PBE} & \multicolumn{2}{c||}{HSE06} &
\multicolumn{2}{c|}{Ga(Si)(Mn)} & \multicolumn{2}{c}{ } &
\multicolumn{2}{c}{PBE} & \multicolumn{2}{c}{HSE06} \\
\hline
\multirow{2}{*}{AlSi-Mn}
&  Ortho &  50 & -356 &  0.48 &  0.84 & 1.195 & 1.353  &
\multirow{2}{*}{GaSi-Mn}                  
&  Ortho & 134 & -169 &  0.599&  0.614&     x &     x  \\
&  \bTet & -48 & -403.0 &  0.435&  0.821& {\bf 2.156} & 2.254  &
&  Tetra &  54 & -249 &  0.370&  0.681& 1.761 & 2.022  \\
\hline
\multirow{2}{*}{AlSi-Tc}
& Ortho  &  76 & -528 &  1.101&  1.174& 1.757 & 1.829  &
\multirow{2}{*}{GaSi-Tc}                  
& Ortho  &  30 & -382 &  0.925&  1.025&     x &     x  \\
& \bTet  & -76 & -604 &  0.485&  1.351& 1.320 & 2.498  &
& \bTet  & -30 & -412 &  0.363&  1.194& 1.181 & 2.223  \\
\hline
\multirow{2}{*}{AlSi-Re}
& Ortho  &  72 & -395 &  0.336&  0.954& 1.050 & 1.614  &
\multirow{2}{*}{GaSi-Re}                  
& Ortho  & 132 & -164 &  0.323&  0.856& 0.902 & 1.483  \\
& Tetra  & -33 & -467 &  0.277&  1.615& 0.862 & {\bf 2.637}  &
& \bTet  & -54 & -296 &  0.135&  1.230& 0.908 & 2.295  \\
\hline
\multirow{2}{*}{AlGe-Mn}
&  \bOrt & -20 & -164 &  0.612&  0.645& 0.000 & 0.000  &
\multirow{2}{*}{GaGe-Mn}                  
&  Ortho &  90 &   26 &  0.504&  0.541&     x &     x  \\
&  Tetra &  20 & -144 &  0.478&  0.638&     x &     x  &
&  Tetra &  68 &    3 &  0.212&  0.489&     x &     x  \\
\hline
\multirow{2}{*}{AlGe-Tc}
& Ortho  &  21 & -402 &  0.754&  0.953&     x &     x  &
\multirow{2}{*}{GaGe-Tc}                  
& Ortho  &  19 & -216 &  0.694&  0.872&     x &     x  \\
& \bTet  & -21 & -423 &  0.311&  1.089& 1.151 & 2.107  &
& \bTet  & -19 & -235 &  0.178&  0.869& 0.880 & 1.816  \\
\hline
\multirow{2}{*}{AlGe-Re}
& Ortho  &  63 & -178 &  0.172&  0.781&     x &     x  &
\multirow{2}{*}{GaGe-Re}                  
& Ortho  &  71 &   40 &  0.199&  0.745&     x &     x  \\
& \iTet  & -43 & -241 &  0.106&  1.415& 0.870 & 2.284  &
& \bTet  & -20 &  -30 &  0.019&  0.861& 0.811 & 1.907  \\
\hline
\hline
\end{tabular}
\caption{\label{tab:data} Stability and bandgap data for various alloy systems.  Element names in parenthesis indicate columns of the periodic table. $\Delta H$ and $\Delta E$ are enthalpy of formation and instability energy in units of meV/atom.  $E_g$ and $E_p$ are the energy gap and pseudogap in units of eV.  PBE and HSE06 are the density functionals.  Newly predicted stable structures are in bold.  Previously known and reconfirmed stable structures are in italics.  Maximum gap and pseudogap are in bold.}
\end{table*}

\begin{figure}
\includegraphics[width=2.75in,angle=-90]{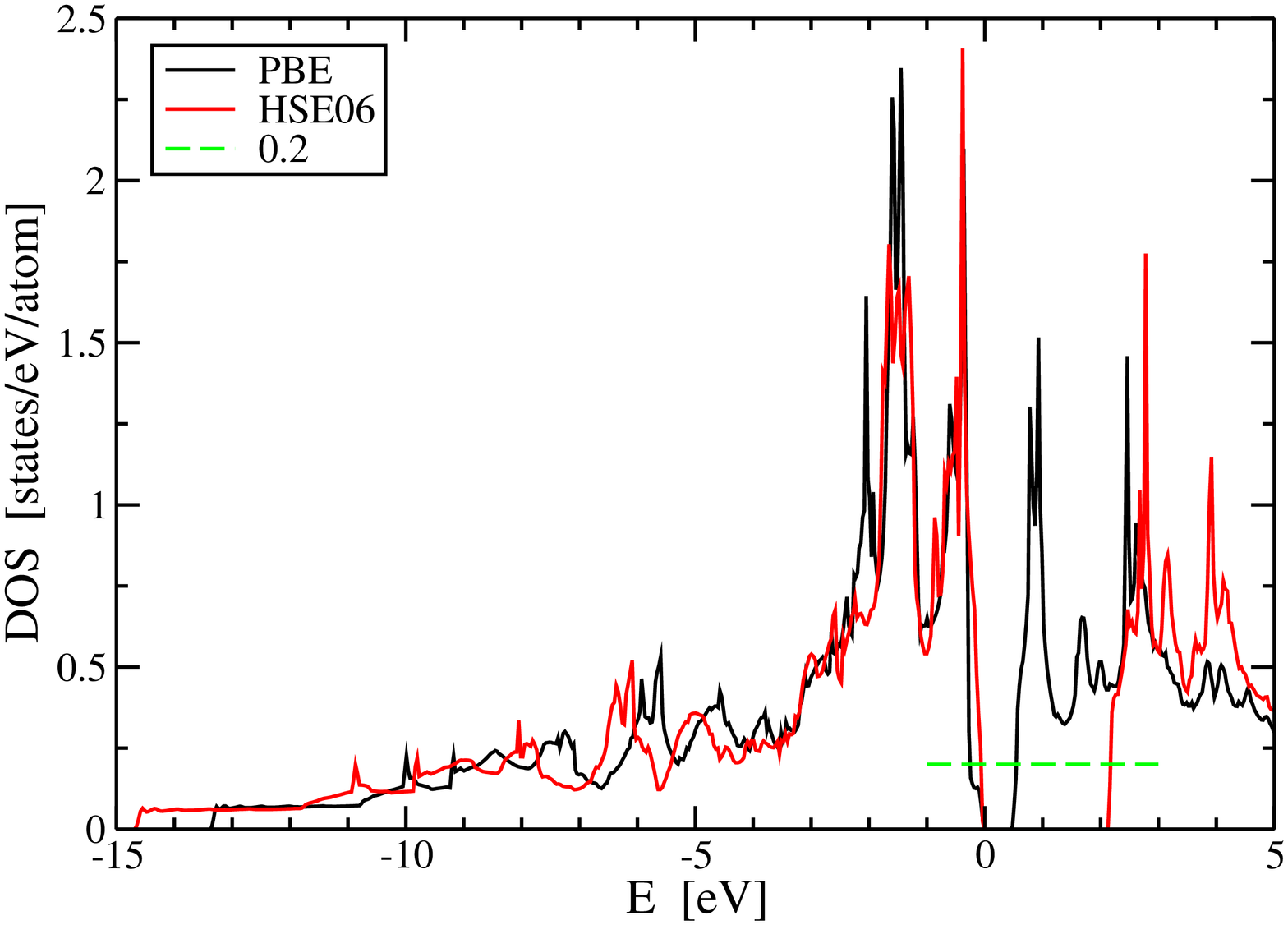}
\includegraphics[width=2.75in,angle=-90]{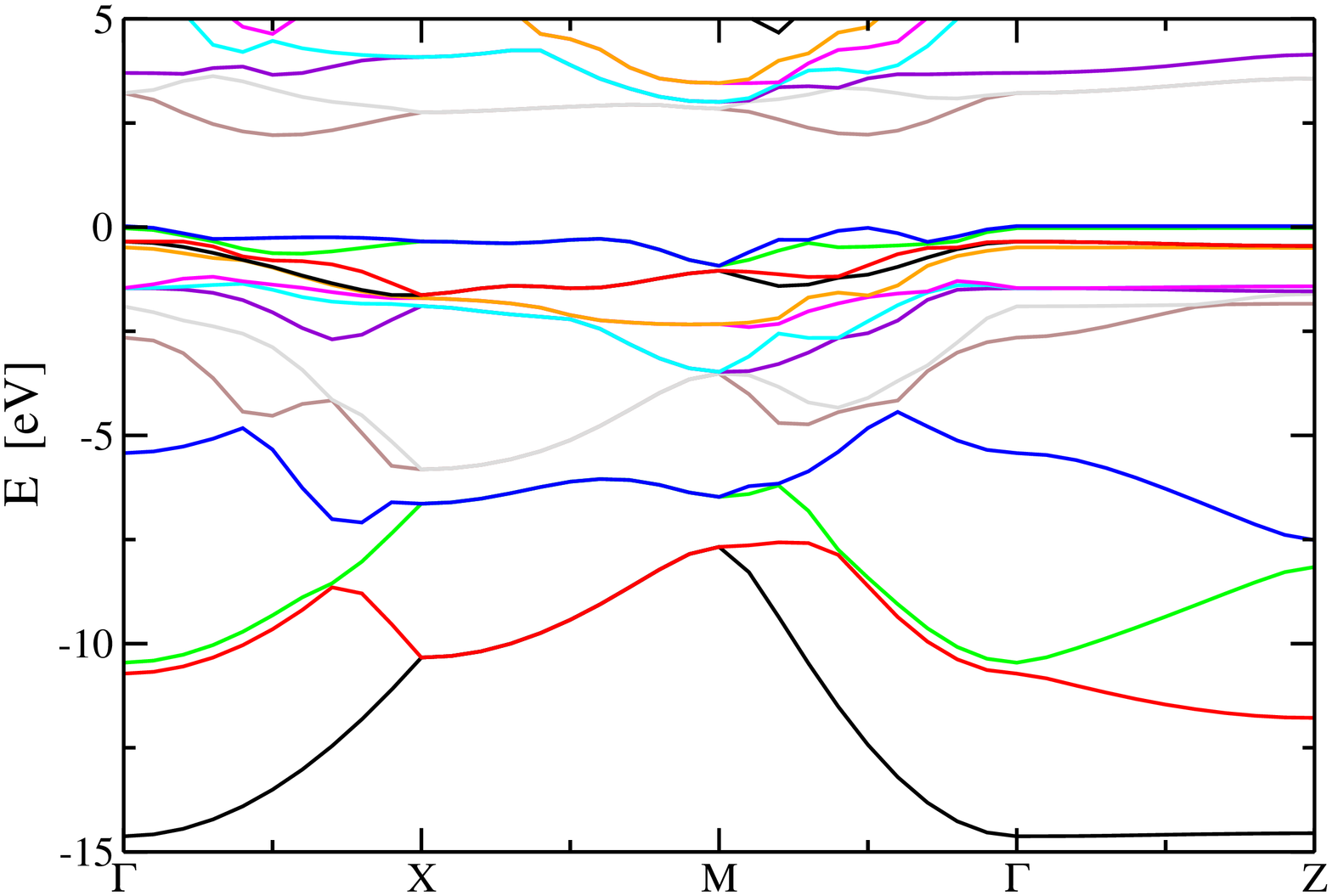}
\caption{\label{fig:dos} (a) Electronic density of states of AlMnSi in the MoSi$_2$.tI6 structure.  Shown are DOS for the PBE and HSE06 functionals with $E_F=0$.  Threshold of 0.2 states/eV/atom defines the pseudogap. (b) Band structure of AlMnSi in the MoSi$_2$.tI6 structure calculated using the HSE06 functional.}
\end{figure}

The electronic density of states and band structure of our predicted AlMnSi structure are shown in Fig.~\ref{fig:dos}.  As expected, the conventional DFT gap is severely underestimated, owing to the derivative discontinuity and delocalization error~\cite{Chan10}.  Inspecting the band structure, we see that the band gap is indirect.  Projecting onto atomic orbitals we find the top of the valence band at $\Gamma$ is purely $d_{xy}$ localized on the Mn atoms, while the bottom of the conduction band, located along $\Gamma-X$ is a hybrid of Al and Si $p_x$ and $p_y$ with Mn $d_{xz}$ and $d_{yz}$ states.  A nearly degenerate conduction band local minimum of similar hybridized character occurs along $M-\Gamma$.  Another interesting feature is the nearly flat valence band minimum along $\Gamma-Z$, which is responsible for the step function-like onset of the density of states near $E=-15$. 

\section{Discussion}

We have predicted a large number of stable intermetallic compounds with hybridization gaps based on chemical ordering on underlying tI6 and oF24 structures, several of which have not been previously reported.  In some cases they may not have been previously reported because the alloy systems have not been throughly evaluated.  In the case of Al-Mn-Si, a chemically disordered phase of structure type oF24 is observed, while we identify the stable state as tP6 (a chemically ordered ternary variant of tI6).  Presumably the entropy of chemical substitution and atomic vibrations stabilizes the disordered structure at high temperature, making identification of the low temperature stable structure difficult.  Preliminary investigation confirms that both configurational and vibrational entropy of oF24 exceed that of tP6.  While the oF24 structure also exhibits a gap, it is smaller than that of tP6, and the chemical disorder can be expected to further diminish the gap in this high temperature phase.  Thus, experimental efforts to identify the true low temperature stable phase are especially welcome.

\acknowledgements
MM and MK were supported by Slovak Research and Development Agency grants APVV-0647-10, APVV-0076-11 and VEGA 2/0111/11.

\bibliography{al4irre}

\end{document}